\documentclass[aps,prl,twocolumn]{revtex4-1}

\usepackage{graphicx}% Include figure files
\usepackage{dcolumn}% Align table columns on decimal point
\usepackage{bm}% bold math
\usepackage{hyperref}

\begin{document}

\title{Topologically protected quantum transport in locally exfoliated bismuth at room temperature} 

\author{C. Sabater$^1$, D. Gos\'albez-Mart\' inez$^1$, J. Fern\'andez-Rossier$^2$, J. G. Rodrigo$^3$, C. Untiedt$^1$, and J. J. Palacios$^3$ } 
\affiliation{\\$^1$Departamento de F\' isica Aplicada, Universidad de Alicante, San Vicente del Raspeig, Alicante 03690, Spain.}
\affiliation{$^2$International Iberian Nanotechnology Laboratory, Av. Mestre Jos\'e Veiga, 4715-330 Braga, Portugal.}
\affiliation{$^3$Departamento de F\' isica de la Materia Condensada, Universidad Aut\'onoma de Madrid, Cantoblanco, Madrid 28049, Spain.}

\begin{abstract}
We report electrical conductance measurements of Bi nanocontacts
created by repeated tip-surface indentation using a scanning tunneling microscope at temperatures of 4 K and 300 K.  
As a function of the elongation of the nanocontact we measure
robust, tens of nanometers long plateaus of conductance $G_0= 2e^2/h$ at room temperature. This observation can be
accounted for by the mechanical exfoliation of a Bi(111) bilayer, a predicted QSH insulator, in the 
retracing process following a tip-surface contact.  The formation of the bilayer
is further supported by the additional observation of conductance steps below $G_0$ before break-up at both temperatures.
Our finding provides the first experimental evidence of the possibility of mechanical exfoliation of Bi bilayers, 
of the existence of the  QSH phase in a two-dimensional crystal, and, most importantly,
of the observation of the QSH phase at room temperature.
\end{abstract}

\pacs{}
\keywords{}
\maketitle
\date{\today}

Topological insulators present a gap in the bulk, but host surface states protected against backscattering by
time reversal symmetry\cite{RevModPhys.82.3045}. This implies that they are immune to non-magnetic disorder-induced localization, i.e.,
they are able to carry electrical current on the surface regardless of imperfections. 2D
TI’s\cite{PhysRevLett.95.226801}, actually predicted before their three-dimensional (3D) 
counterparts\cite{Zhang09-1}, are expected to exhibit the
so-called quantum spin Hall (QSH) phase, a spin filtered version of the integer quantum Hall effect\cite{QHE}. 
While the most exotic experimental manifestation of this phase is through a nearly 
universal spin Hall conductivity of $\approx e/2\pi$, a
truly universal charge transport is expected to manifest, e.g., as a two-terminal conductance $G_0 = 2e^2/h$.

To date, two types of 2D systems have been
predicted to be QSH insulators: two-dimensional crystals such as graphene\cite{PhysRevLett.95.226801} or 
Bi(111) bilayers\cite{PhysRevLett.97.236805} and
semiconductor heterojunctions such as CdTe/HgTe\cite{König02112007} or, more recently, InAs/GaSb
quantum wells\cite{PhysRevLett.100.236601}.
Transport measurements in CdTe/HgTe\cite{König02112007} and InAs/GaSb\cite{PhysRevLett.107.136603} 
quantum wells have revealed the presence of protected edge states and provided the first
experimental evidence of the QSH phase to date. 
The QSH state in 2D crystals, on the other hand, has not been experimentally confirmed to date.
The fact that spin-orbit coupling (SOC) in graphene is so weak precludes the observation of the QSH phase in this material. 
Bismuth, on the contrary, presents a naturally large SOC, its mechanical and electronic properties are well
characterized for bulk and surface\cite{Hofmann06}, in nanowire form\cite{cm9811545},
and, recently, the existence of edge states in Bi(111) bilayers has been reported\cite{PhysRevLett.109.016801}. 
Other proposals stay, at this moment, at a more speculative level\cite{PhysRevLett.107.076802}.

\begin{figure}[tb]
  \begin{center}
\includegraphics[width=8 cm]{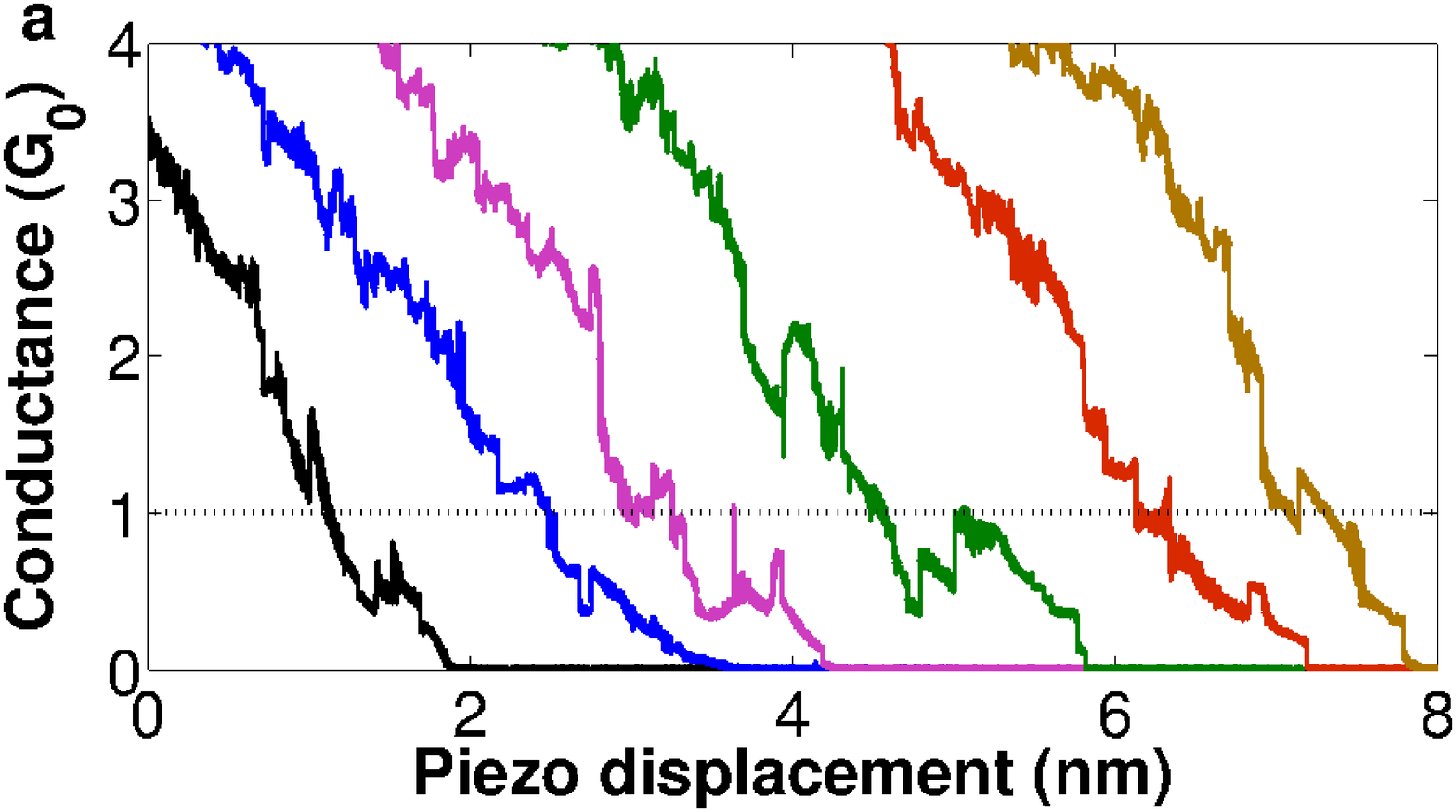}\\
\includegraphics[width=8 cm ]{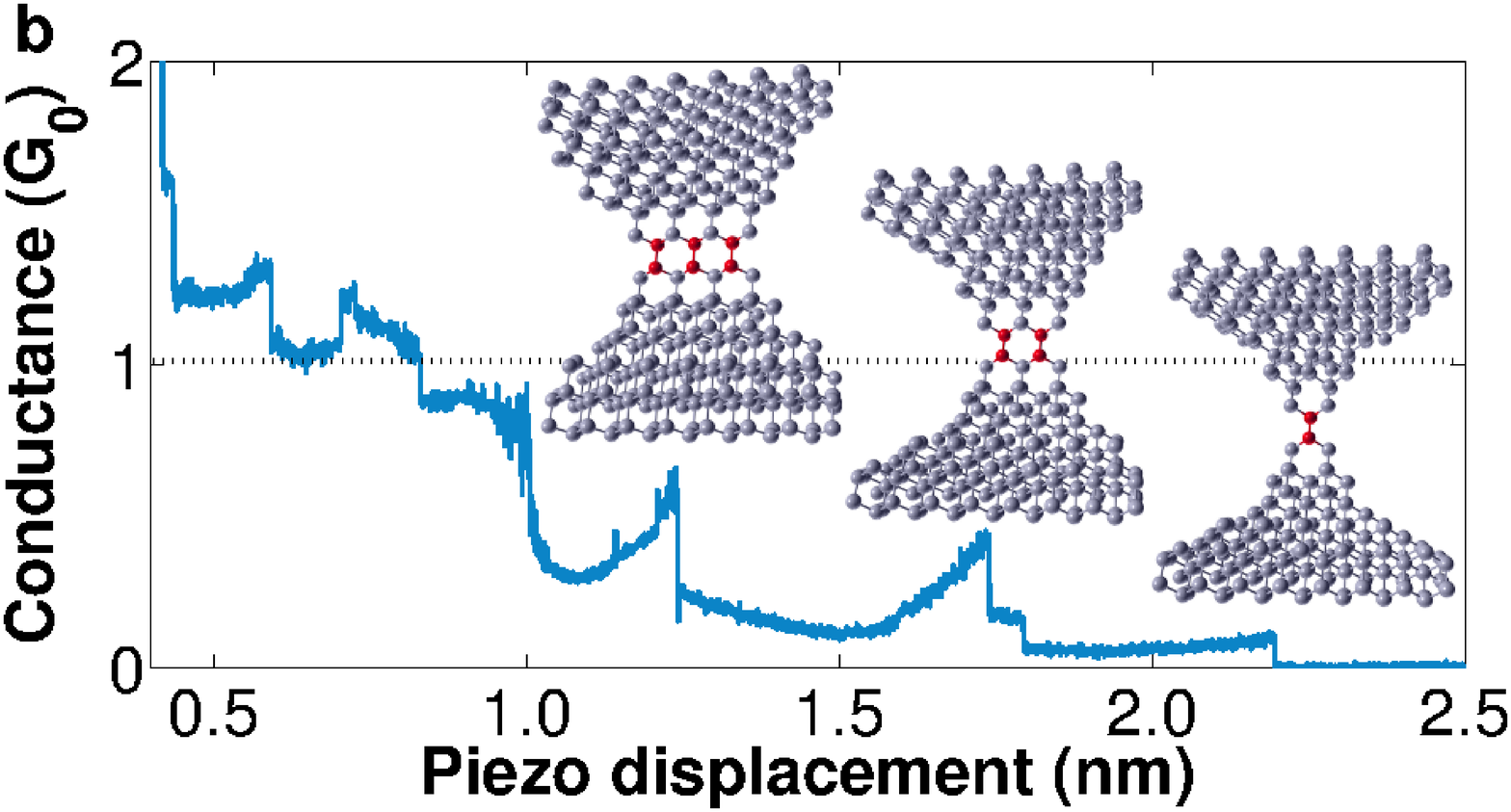}
  \caption{(a) Selected curves obtained during the process of rupture of a Bi
nanocontact formed after indentation of a tip on a surface at 4 K. 
(b) Detail of a selected curve where four conductance sub-plateaus can  be appreciated below $G_0$. The
insets represent possible atomic configurations compatible with the sub-plateau values. }
  \end{center}
\end{figure}

Cleavage techniques are becoming common in the quest for 2D crystals\cite{Radisavljevi11}. With a few 
exceptions\cite{Hong10}, they remain largely unexplored in the field of topological insulators (TI's).
We report here, using  scanning tunneling microscope (STM) based  mechanical and electrical characterization techniques,
the first evidence of the QSH phase in a two-dimensional crystal such as an exfoliated Bi(111) bilayer. 
The procedure follows closely that of creation of atomic size metallic contacts\cite{Agrait:physrep:03}, namely, 
repeated indentations of a tip against a surface while measuring the current at low voltages (typically 100 meV). 
The resulting conductance traces
contain information on the atomic and electronic structure of the contact.
The experiments where done at room temperature (and ambient conditions) 
and at low temperatures (4K) in cryogenic vacuum using two Bi samples of different origin\cite{supplementary}.
Related experiments have been reported in the past\cite{Krans94,PhysRevLett.88.246801,PhysRevLett.78.4990},
lacking a convincing interpretation of the results, but have, in part, motivated our work.

As a rule of thumb (in particular for monovalent elements) each atom forming the minimum cross section of a metallic 
nanocontact contributes to the conductance with a value in the vicinity of $G_0$\cite{Agrait:physrep:03}. 
Thus, on retracing the tip after indentation, 
one-atom cross-section contacts formed right before complete break-up are signaled by a conductance plateau (as a function
of the STM piezo elongation) near the quantum of conductance $G_0$.  
Unlike their more common metallic counterparts, the conductance 
traces of Bi nanocontacts at 4 K present several sub-plateaus below $G_0$
before break-up (see Fig. 1a).  In Fig. 1b, we have singled out one trace that manifestly shows 
four sub-plateaus before the final rupture. Since each plateau corresponds to an elastic
deformation of an atomic configuration in between plastic deformation events, one can only conclude that few-atom cross
section Bi nanocontacts (as the ones shown in the insets) are much less conducting than their metallic counterparts. 
Being a common practice to record conductance histograms to statistically characterize nanocontacts\cite{Agrait:physrep:03},
we have done so using thousands of rupture traces. We find a large statistical weight below $G_0$, 
but no characteristic values can be appreciated\cite{supplementary}.

\begin{figure}
\begin{center}
\includegraphics[width=8 cm]{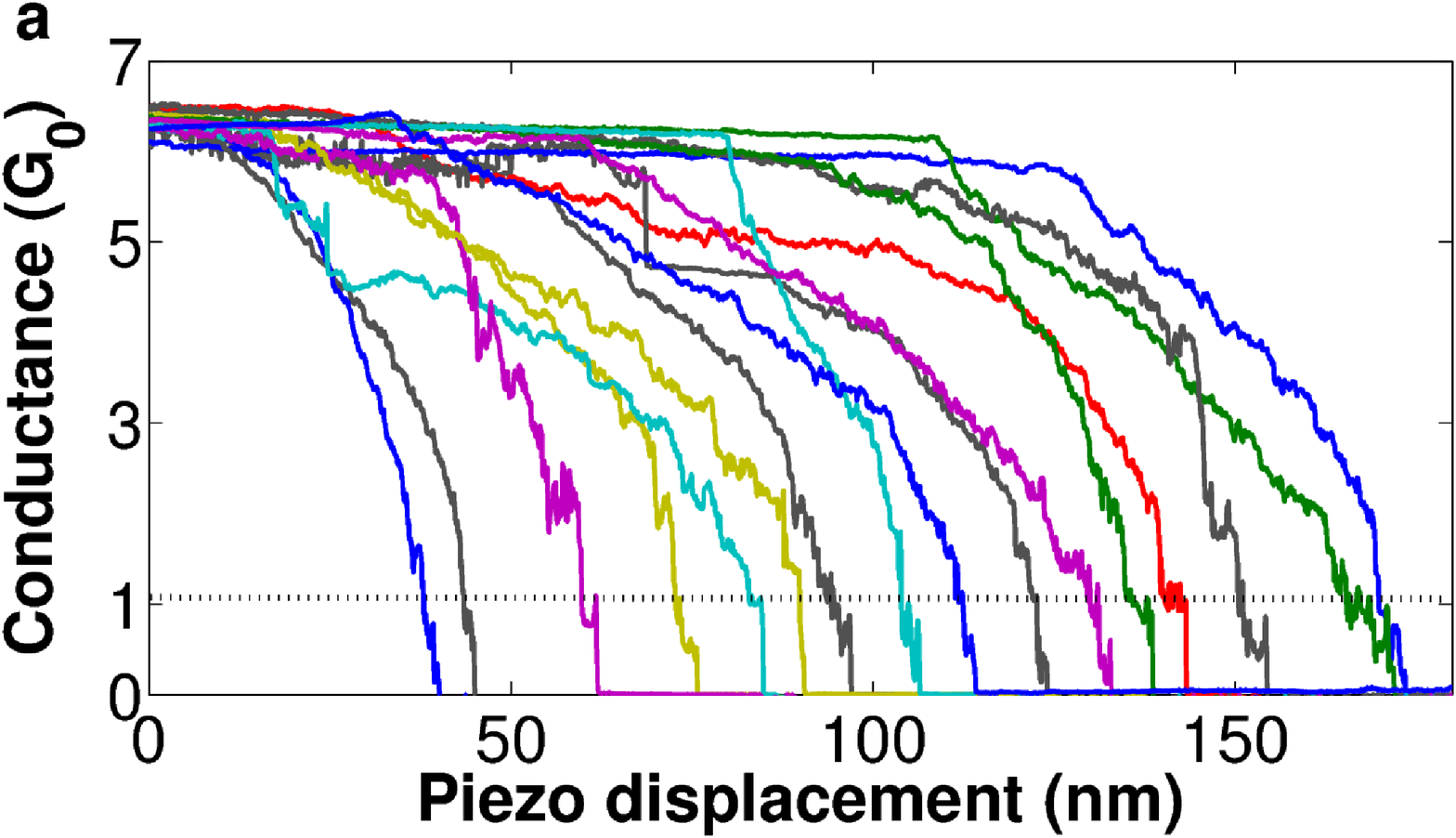} \\
\includegraphics[width=8 cm]{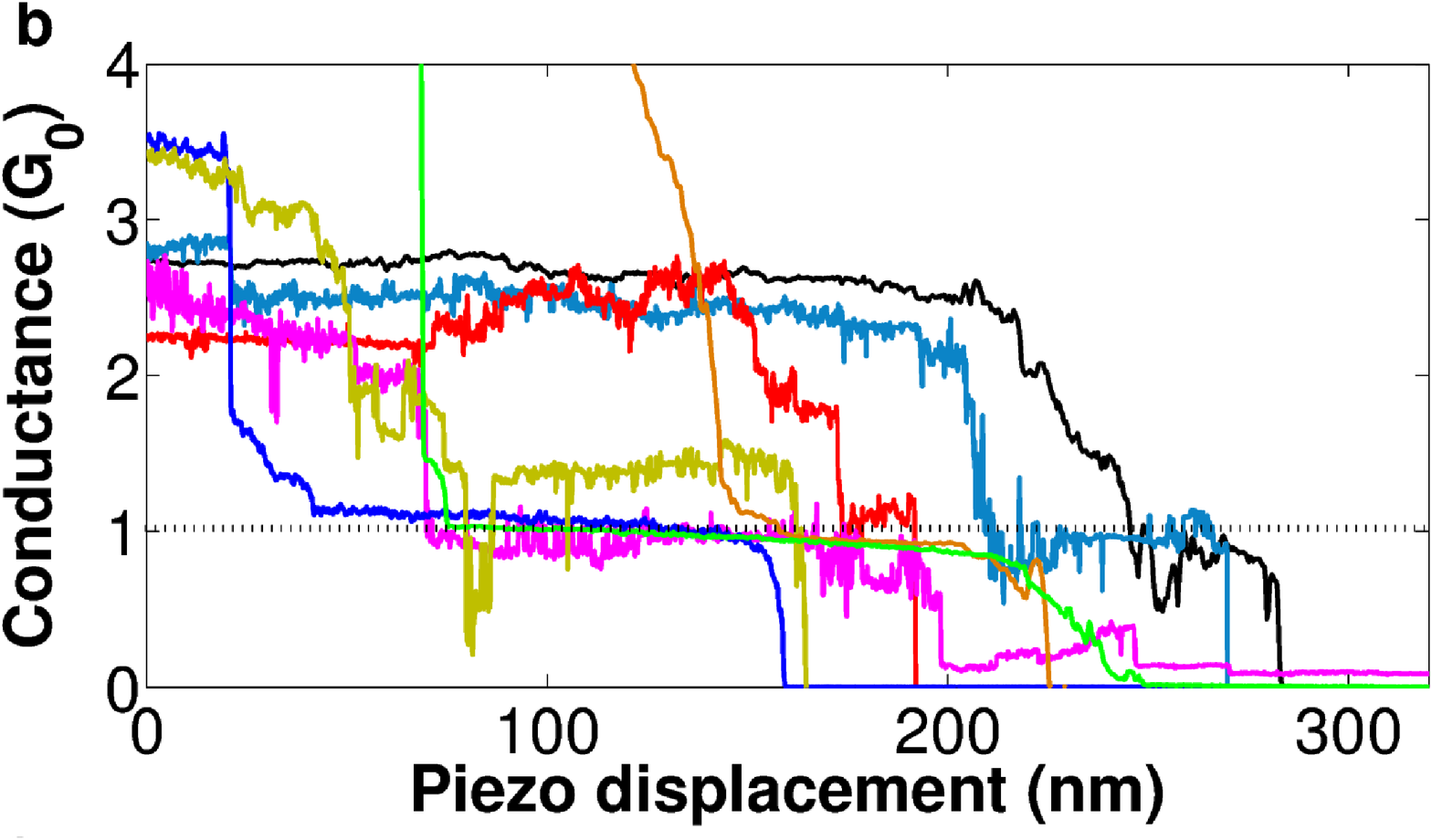} 
\includegraphics[width=8 cm]{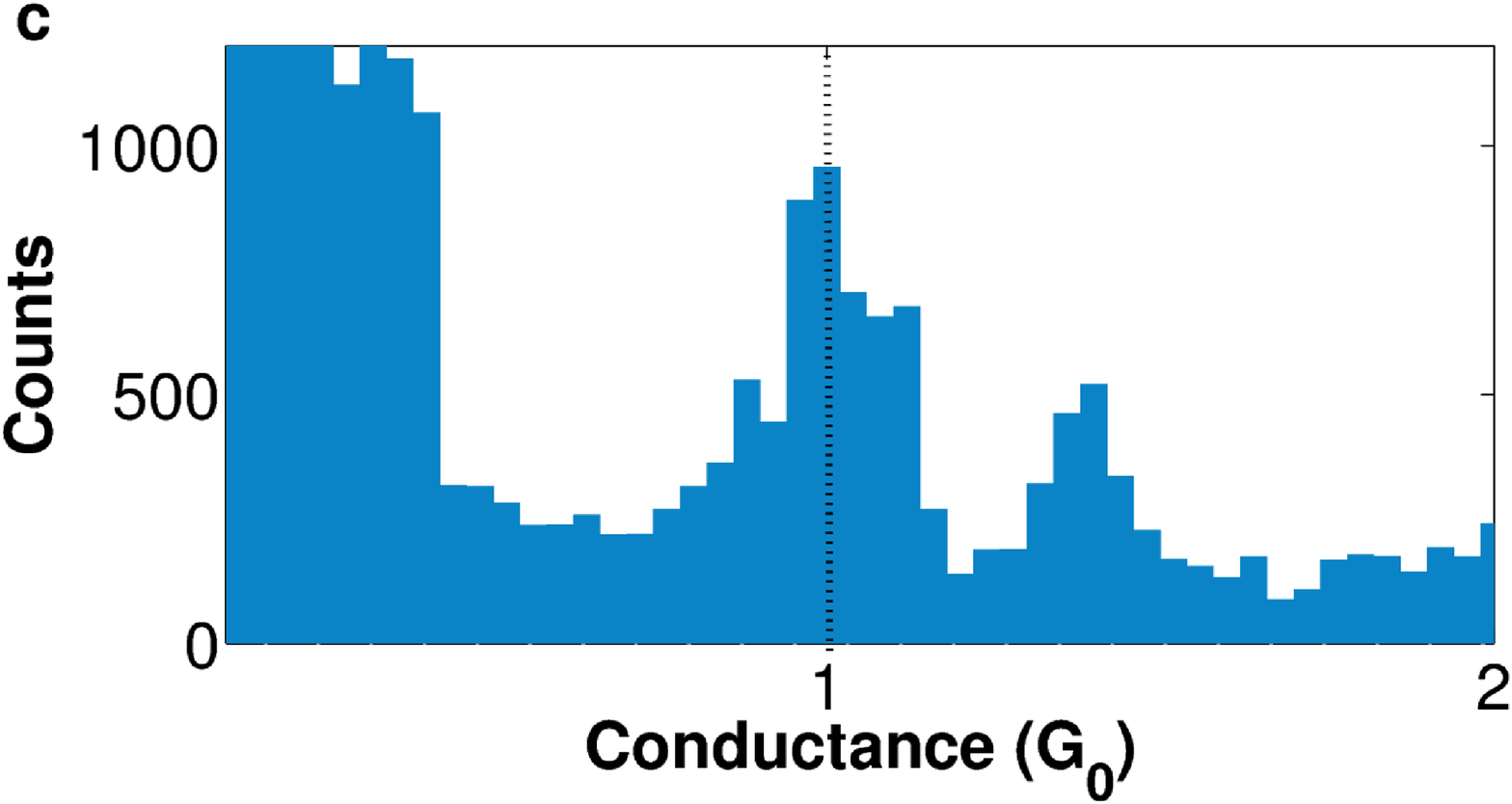} 
  \caption{ (a) Examples of conductance recorded during the process of rupture of a Bi nanocontact formed by 
repeated indentation of a Bi tip on a Bi surface at 300 K. 
%Note the large piezo displacement (in the range of 100 nm) 
%required to produce  conductance variations similar to the ones obtained at 4 K (1 nm), shown in Fig. 1. 
%The behavior of the conductance curves in the initial stages of the elongation process (i.e., at high conductance) 
%denotes  an unexpectedly large elasticity of these Bi nanocontacts.  After the initial elastic process, 
%plastic deformations occur and conductance features and short plateaus appear at several values, including $G_0$.
(b) Conductance traces exhibiting 
plateaus with elongations of tens and hundreds of nanometers attributed to the occasional exfoliation 
of large flakes of Bi(111) bilayers between tip and substrate. Most of the plateaus are pinned at $G_0$. 
(c) Histogram of conductance obtained from tens of traces exhibiting long plateaus. The histogram
exhibits a clear peak at $G_0$ which reveals that this value is particularly frequent and robust. }
  \end{center}
\end{figure}

In air at 300 K the mechanical stability of the experimental set-up is inferior to that at low temperatures, making
it more difficult to carry out measurements in a systematic way. Previously, the samples
were fully characterized\cite{supplementary}, paying attention to surface
contamination and, in particular, to oxidation. This turned out to be very small, as long known for Bi\cite{Tahboub79}.
Starting from conductance values at a maximum indentation of the order of 10$G_0$, most of
the traces either do not break for the 
allowed range of the piezo, typically up to 300 nm, or fall to zero showing no apparent 
reproducible behavior. Starting from smaller indentations (up to 3-6 $G_0$), however, 
consecutive traces may repeatedly show small features or short plateaus, 
including plateaus at $G_0$ (see Fig. 2a) which were
completely absent in the 4 K measurements. Notice also the different length scales in Figs. 1a and 1b (1 nm) and 
Figs. 2a and 2b (100 nm).
Most notably, long conductance plateaus appear on occasions on some traces (see Fig. 2b), 
the length of these reaching up to hundreds of nanometers. While some of them correspond to the 
initial elongation elastic process (the ones around $3G_0$), most of them appear after a few plastic events.
One way to quantify the information in these traces is to plot a conductance 
histogram only with them (we have selected those exhibiting plateaus
longer that 50 nm regardless of the plateaus conductance values). The result is incontrovertible, 
showing a large peak at $G_0$ since most of the plateaus are largely pinned at the quantum of conductance (see Fig. 2c).

An exact conductance value of $G_0$ is generally expected to occur when, first, the Fermi wavelength is  comparable to
the constriction width and, second, the constriction potential is adiabatic in the 
current direction.  This happens, for instance, in a gated 
two-dimensional electron gas\cite{van91} or in monovalent metal atomic contacts\cite{Agrait:physrep:03}. 
Crystalline bulk Bi presents two types of carriers (electrons and holes) with long Fermi wavelengths
(tens of nm) and unusually small masses\cite{PhysRevB.52.1566}. 
A first scenario,  already put forward to explain the early  experiments in Ref. 
\onlinecite{PhysRevLett.88.246801}, invokes the formation of large cross section three dimensional constrictions 
(thousands of atoms) so that the lateral movement of the long-wavelength electrons gets quantized. 
It is, however, difficult to predict how the bulk electronic structure can carry over to 
constrictions presumably lacking long crystalline order. In addition,  adiabaticity 
imposes severe restrictions to the shape of the constrictions. Furthermore, it is quite hard to imagine 
how these constrictions may be stable during very long piezo displacements. Even more difficult to accept 
is the fact that they must exist right before rupture without then going through a very large number of plastic 
deformation events. A second scenario where atom-size constrictions are responsible 
for the appearance of $G_0$ is ruled out by the observation of sub-quantum plateaus
in most traces (see Figs. 1 and 2) and also by our calculations (see below).

We now give our rationale to our findings. 
Bulk Bi can be viewed as a layered material in the (111) direction\cite{supplementary}; much as graphite,
but with a stronger electronic coupling between layers. In fact, a Bi(111)
bilayer is similar to graphene in many respects, in particular the atomic structure is that of a puckered honeycomb lattice,
with the atoms of one sublattice shifted perpendicularly with respect to those of the other sublattice.
A Bi(111) bilayer has been predicted to be a QSH insulator\cite{PhysRevLett.97.236805} with a large bulk gap
and three pairs of helical edge states\cite{Wada11}. This odd number of edge states guarantees
that the transport properties of a Bi(111) flake should be robust against size, shape, and weak disorder. In particular, 
it should generically exhibit a universal two-terminal conductance of $G_0$.

\begin{figure}
  \begin{center}
\includegraphics[width=8 cm]{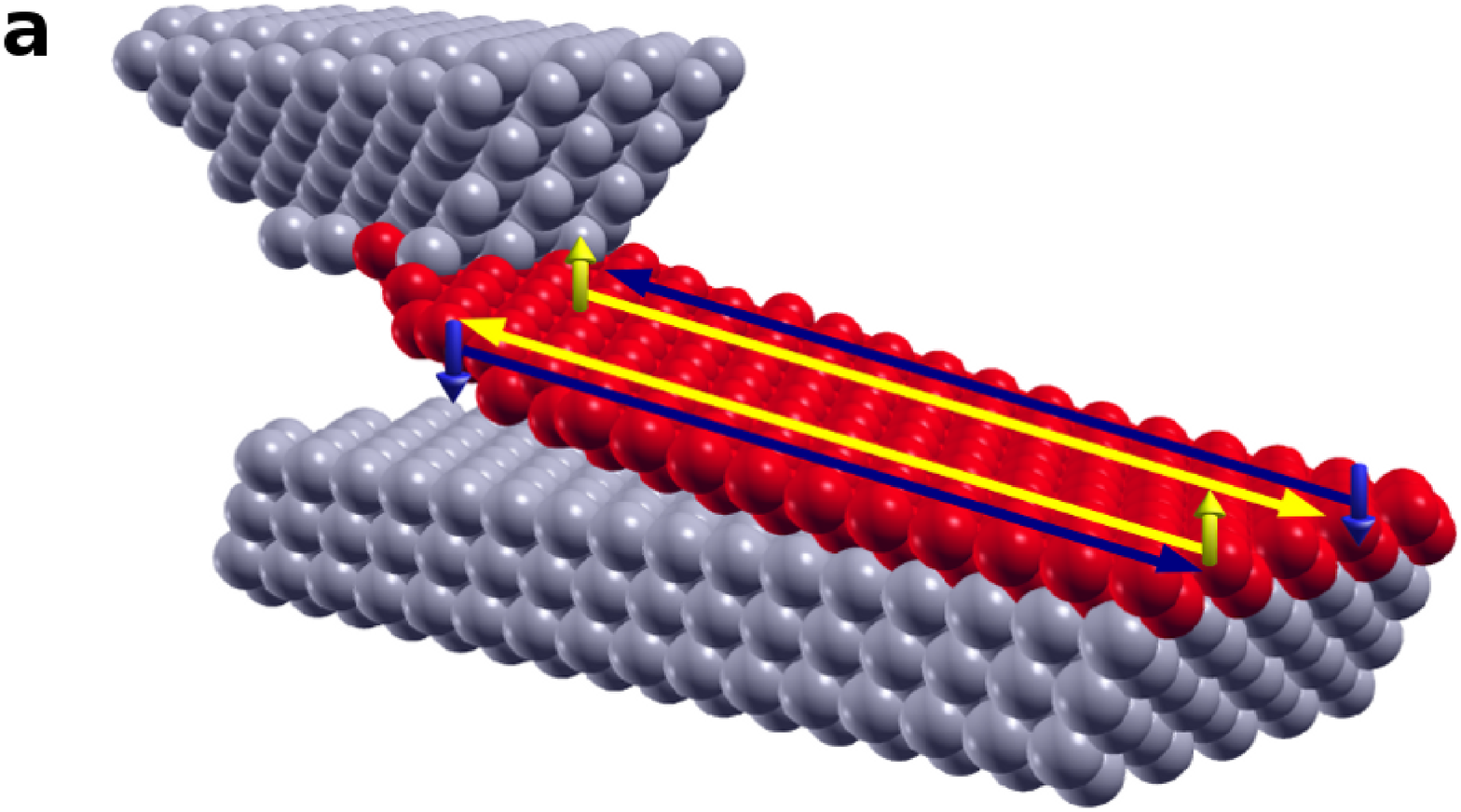} \\ 
\includegraphics[width=8 cm]{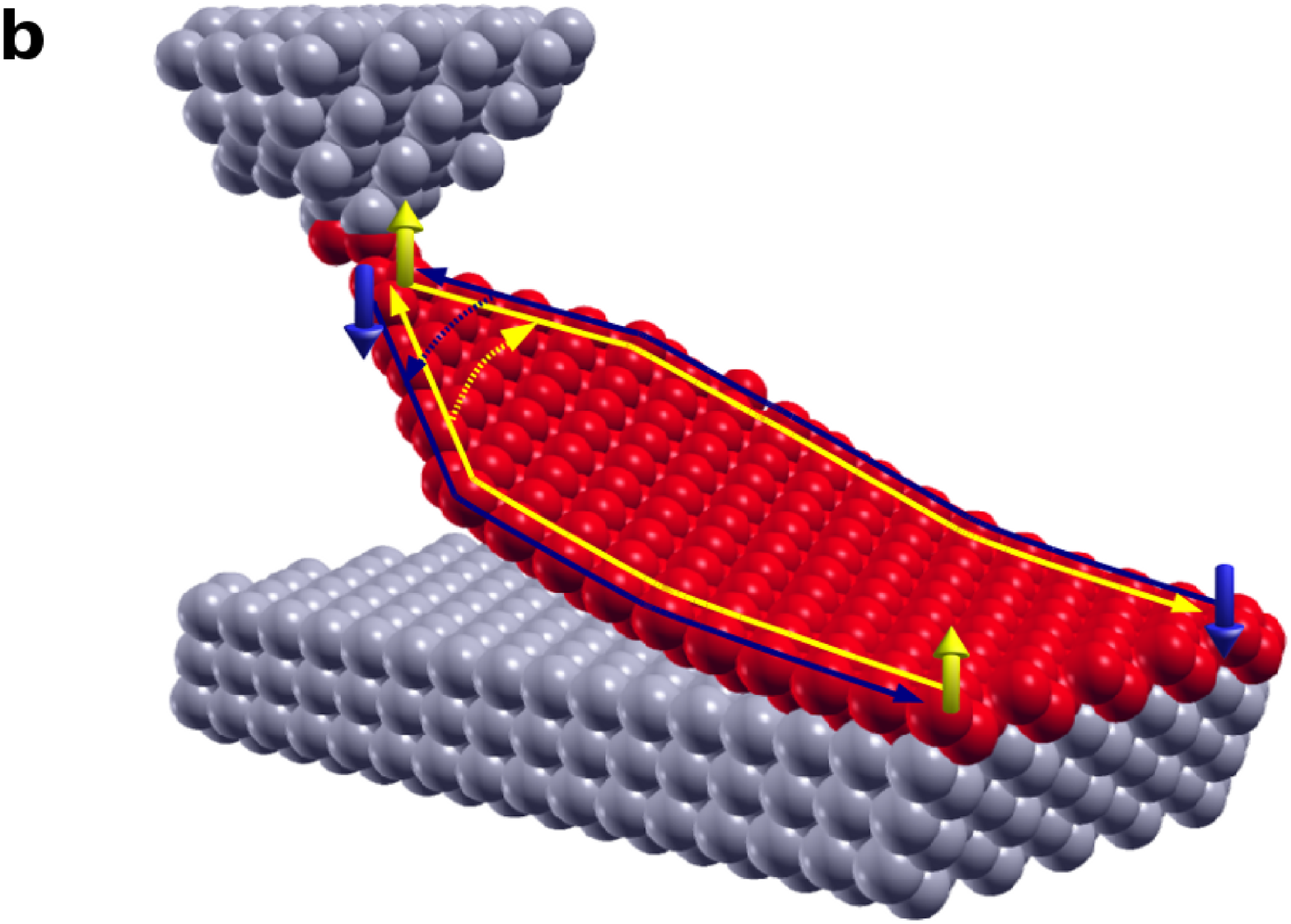} \\
\includegraphics[width=8 cm]{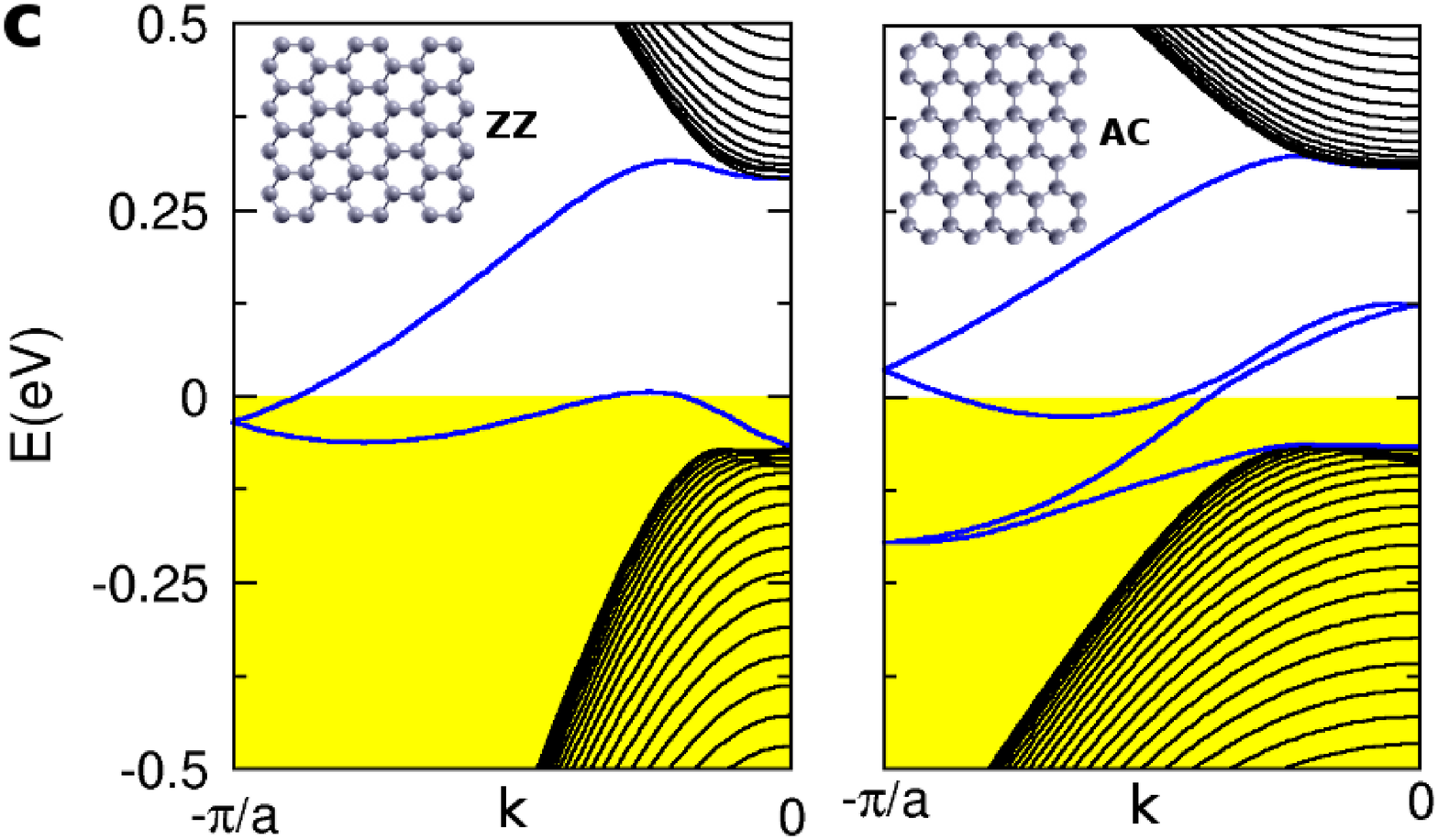}
\caption{(a) Pictorial representation of the proposed process of exfoliation of a Bi(111) bilayer (in red)
after contact with the STM tip. One (out of three) 
helical edge channel is also represented. (b) Same as in (a) but for the last stages of piezo elongation 
before break-up where a few-atom constriction forms. (c) 
Band structure of Bi(111) bilayer nanoribbons along ZZ (left) and AC 
(right) directions as obtained from a self-consistent tight-binding model. 
The insets show a top-view section of these nanoribbons.  The edge bands connecting valence and conduction bands are shown 
in blue. The Fermi energy (set to zero) crosses three times these edge bands which gives rise to three 
conduction channels. }
  \end{center}
\end{figure}

Given the layered structure of Bi, the general picture that naturally emerges from our experiments is that of a
contact-induced exfoliation of a Bi(111) bilayer (see Figs. 3a and 3b). 
We first note that the inter-bilayer coupling is about ten times weaker than
the intra-bilayer covalent bonding\cite{Hofmann06}.  The Bi(111) surface presents a Debye temperature in the 70-80 K 
range\cite{Mayrhofer-Re12} so that at ambient conditions
it should not come as a surprise that, once a gentle contact is made on the appropriate surface orientation, 
retracing the tip can peel off a single
bilayer. One extreme of this bilayer (while the rest remains in contact with the surface) sticks to 
the tip which now acts as an electrode (see Figs. 3a, 3b and Supplemental Material movie\cite{supplementary}). 
The specifics of the proposed mechanical exfoliation are unknown to us, being unclear whether
the tip breaks the surface or simply elastically deforms it after indentation, 
and whether slip directions on the (111) planes favor slipping over detaching mechanisms. 
The number of atoms participating in the contact at maximum conductance is
also unknown, but it remains large enough after exfoliation so that the theoretically expected intrinsic
conductance $G_0$ of the bilayer is not masked by the contact resistance. 

With this picture in mind, the conductance sub-plateaus below $G_0$ can only reflect the breaking of the tip-bilayer
contact or the progressive breaking of the bilayer, a process in which a 
bidimensional nanocontact may form (see Fig. 3b and insets in Fig. 4a). 
Since a disordered Bi(111) bilayer is expected to exhibit
a maximum conductance of $G_0$, we can only expect the nanocontacts thus formed to conduct less due to inter-edge
backscattering at the constriction (this is more clearly observed at 4 K).  Noticeably, at 4 K, long plateaus 
do not appear near $G_0$. 
We attribute this absence to the stronger inter-bilayer binding at low temperatures which prevents the 
exfoliation of a sufficiently large flake.  Interestingly, the up-turns of the sub-plateaus rarely surpass $G_0$.  

We now support our hypothesis with atomistic calculations.
We model Bi with a 4-orbital tight-binding Hamiltonian with the parametrization given by
Liu and Allen\cite{PhysRevB.52.1566}. In order to account properly for the electronic structure at the edges, 
we carry out self-consistent calculations
including local electron-electron interactions\cite{supplementary}. 
As can be seen in Fig. 3c, Bi(111) bilayer nanoribbons grown along zigzag (ZZ) and armchair (AC)
crystallographic orientations present a gap with valence and conduction bands connected in a continuous manner
by an edge state band. This edge band cuts three times the Fermi energy and 
presents the same topology as that computed with
more sophisticated DFT calculations\cite{Wada11}, which gives us confidence in our model. We have also
verified that the nature of the wave functions in this band follows that expected in a 
QSH insulator\cite{supplementary}.

\begin{figure}
  \begin{center}
\includegraphics[width=8 cm]{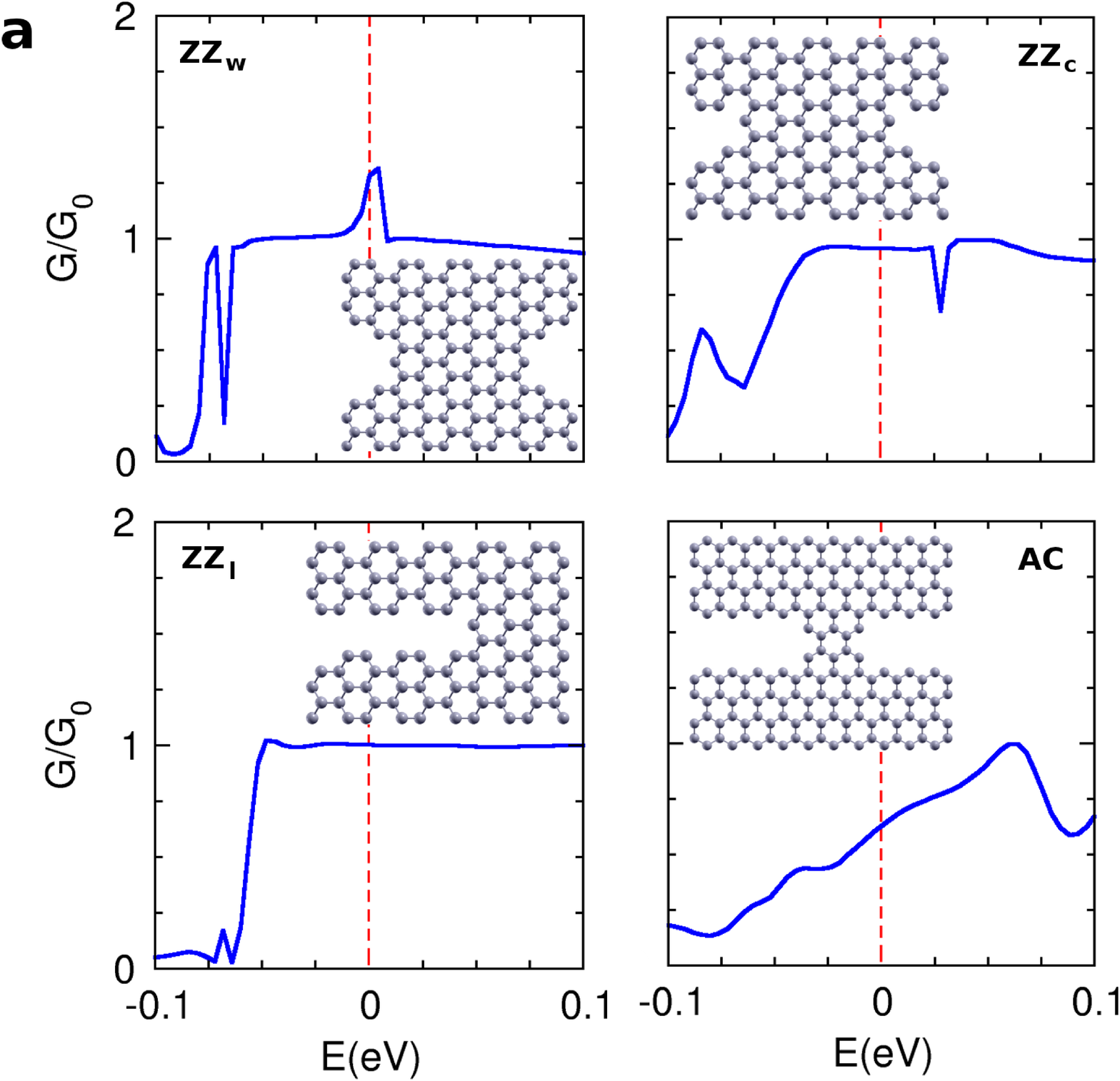} \\
\includegraphics[width=8 cm]{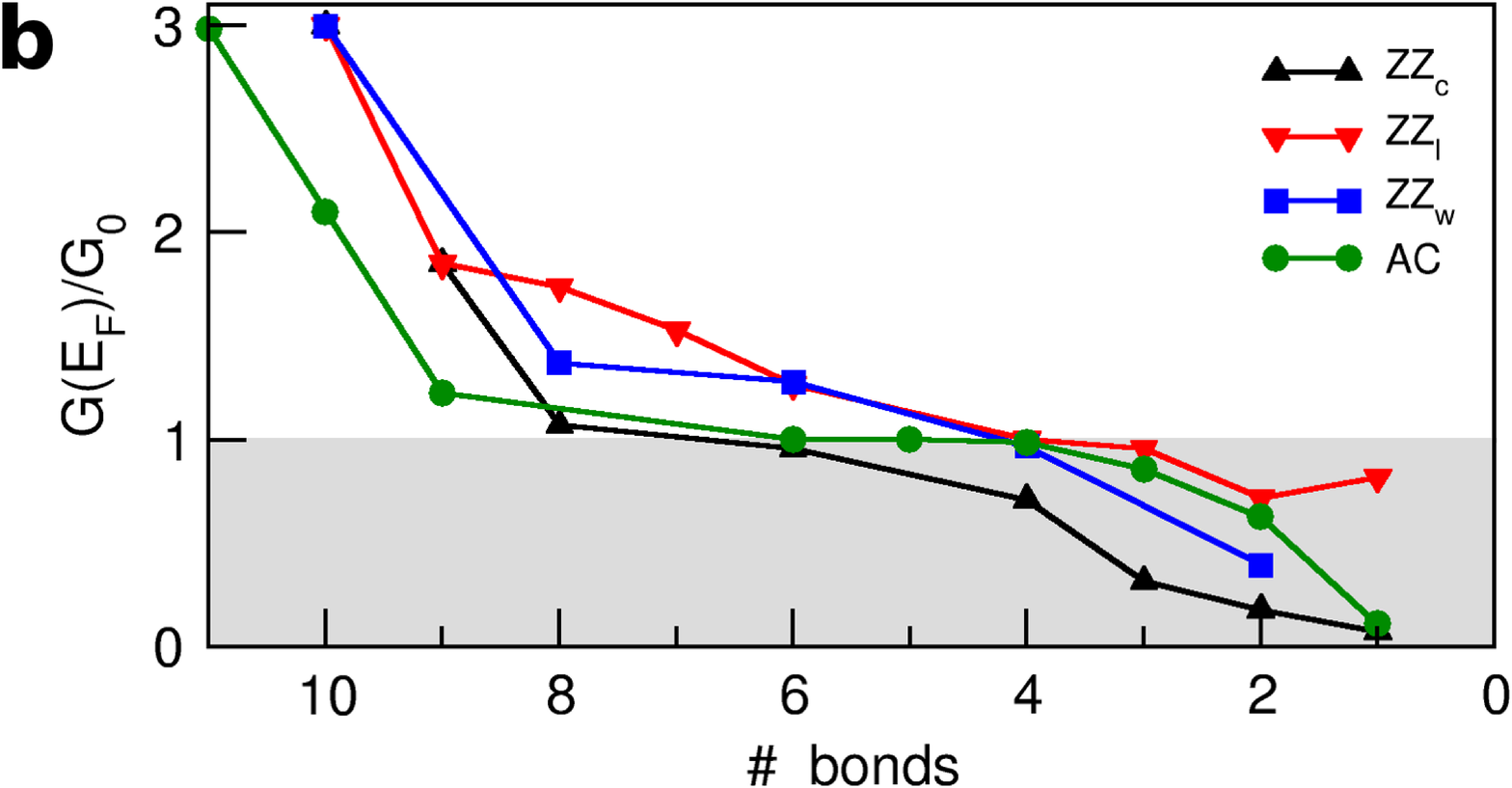}
  \caption{ (a) Conductance vs. electron energy for four constriction models in 
ZZ and AC Bi(111) bilayer nanoribbons. (b) 
Conductance at the Fermi energy for many cases as those in (a) as a function of the number
of bonds remaining in the constriction. The sub indexes denote central- (c), lateral- (l), and wedge-type (w)
constrictions.}
\label{transport}
  \end{center}
\end{figure}

Equipped with a reliable atomistic description of Bi, we now perform coherent quantum transport calculations.
(While inelastic effects cannot be entirely ruled out, bulk 
Bi carriers are known to have very long mean free paths even at room temperature\cite{Das87}.)
We introduce constrictions in the nanoribbons emulating the breaking of the bilayer flake (see insets in 
Fig. 4a).  For simplicity's sake we have chosen to perform this type of calculation instead of the one 
for a contact between a tip (of unknown shape) and the flake (as shown in Figs. 3a and 3b). 
A common way to calculate the conductance (as a function of energy) $G(E)$ 
is through the Green's function formalism and the partitioning technique\cite{ALACANT,Jacob11,supplementary}, now
taking fully into account the atomic SOC\cite{Gosalbez-Mar11,supplementary}. Results of $G(E)$
for four different constriction geometries in AC and ZZ nanoribbons are shown in Fig. 4a. 
Figure 4b shows $G$ at the Fermi energy ($E_F=0$) for these four 
geometries as a function of the number of bonds at the constriction (i.e., as a function of the width).
Only in disorder-free ribbons with perfectly defined edges the three channels can fully transmit. 
For constriction widths just two or three bonds smaller than that of the ribbons
the conductance drops in all cases to values in the close vicinity of $G_0$.
Only constrictions featuring less than approximately five bonds conduct below $G_0$. 
(The robust appearance of $G_0$ also holds for other types of disorder\cite{supplementary}.) 
These results can be simply understood in terms of strong intra-edge backscattering of two of the edge channels 
and absent or partial inter-edge 
backscattering of the remaining helical pair at the narrowest section of the constriction (see Figs. 3a and 3b). 
Our estimate of a minimum of five bonds for a constriction to support
$G(E_F)= G_0$ is consistent with the maximum number of plastic events typically seen in the experiments below $G_0$
before the constriction definitely breaks (see, e.g., Fig. 1b). 
 
In summary, we have reported and 
offered a consistent interpretation to the appearance of extremely long quantum of conductance plateaus in the
breaking process of Bi 
nanocontacts. We attribute it to the occasional local exfoliation of Bi bilayers, predicted to be 2D topological insulators.
Other interpretations cannot be ruled out, but should account for three extraordinary facts:
(i) the quantum of conductance cannot be associated with a single-atom contact, but with a nanoscopic constriction,
 (ii) it appears at room temperature, and (iii) it is robust to small concentration of 
contaminants since the experiments are performed in air.

The first two authors, CS and DGM, contributed equally to this work. The former on the experimental side
and the latter on the theoretical part.
We thank the Spanish MICINN for financial support through Grant Nos.  FIS2010-21883, FIS2011-23488, and CONSOLIDER
CSD2007-0010 and Generalitat Valenciana through Grant Nos. ACCOMP/2012/127 and PROMETEO/2012/011.  
DG acknowledges the Centro de Computaci\'on Cient\' ifica at UAM for computational support.
Cristina Almansa is gratefully acknowledged for her help with the transmission electron microscope.

\end{document}